\documentclass[12pt]{iopart}

\begin{document}

\input epsf
\jl{4}

\title[E864 Negative Strangelet Search]{Negatively
Charged Strangelet Search using the E864 Spectrometer at the AGS}

\author[G.\ Van Buren]{Gene Van Buren\dag~for the E864 Collaboration\ddag}
\address{\dag~MIT 26-403, 77 Massachusetts Avenue, Cambridge, MA 02139-4307,
USA}
\address{\ddag~Univ~Bari--BNL--UCLA--UC~Riverside--Iowa~State--Univ~Mass--MIT--
Penn~State--Purdue--Vanderbilt--Wayne~State--Yale}


\begin{abstract}
We provide a status report on the progress of searching for negatively
charged strangelets using the E864 spectrometer at the AGS. About 200
million recorded events representing approximately 14 billion 10\%
central interactions of Au~+~Pt at 11.5 GeV/c taken during the 1996-1997
run of the experiment are used in the analysis. No strangelet candidates
are seen for charges $Z=-1$ and $Z=-2$, corresponding to a 90\% confidence
level for upper limits of strangelet production
of ${\sim}1 \times 10^{-8}$ and ${\sim}4 \times 10^{-9}$ per central
collision respectively. The
limits are nearly uniform over a wide range of masses and are valid
only for strangelets which are stable or have lifetimes greater than
$\sim$50 ns.
\end{abstract}

\section{Introduction}

The discovery of quarks nearly 30 years ago solidified the understanding
of hadrons as multi-quark systems. To this day, no clear evidence
has been found for the existence of larger quark systems
(with more than three quarks).
However, theoretical calculations [1,2,3]
using the MIT Bag Model of quark confinement [4] predict the possibility
that such quark systems with roughly equal numbers of $u$, $d$, and $s$ quarks
in color-singlet states, or
Strange Quark Matter (SQM), might be metastable, stable, or even the true
ground state of hadronic matter.

Because stability of SQM mandates similar quantities of $u$, $d$, and $s$
quarks, such systems are expected to carry only a light overall charge,
if any, despite high mass.
Searches for small SQM systems, termed $strangelets$, have relied upon this
property of low charge-to-mass ratios. Recent calculations have even shown
that negatively charged strangelets may be the most favorable for
stability [1].

Strangelets
have been postulated to be producible in the baryon-rich environment of
heavy ion collisions [5]. Mechanisms for strangelet
production include the coalescence
model [6], thermal model [7], and Quark-Gluon Plasma (QGP) distillation [8,9].
The coalescence and thermal models rely on the production of hyperfragments,
which may then decay into more stable strangelet states. This scenario
is difficult to achieve for negatively charged strangelets
in heavy ion collisions because of the extremely rare production of large,
negatively charged or neutral hyperfragments.

In QGP distillation, large numbers of $s{\bar s}$ quark pairs
are believed to be abundantly produced inside the QGP. Subsequently,
${\bar s}$ quarks
are evaporated more readily by forming kaons with the copiously available
$u$ and $d$
quarks, leaving a QGP which is strangeness-rich. Such an environment is
ideal for strangelet formation. However, even QGP formation has not
been indisputably observed [10]. That strangelets may be distilled is
even more speculative [11]. Observing
negatively charged strangelets in such collisions would therefore be
strong evidence for the formation of QGP.

A status report of the search for negatively charged strangelets with the
E864 apparatus is presented in this contribution. The expected limits
from this search are shown and compared with previous experiments. The
implications for QGP formation are also discussed and presented.

\section{The Apparatus}

The E864 experiment
is devised specifically to search for low charge to mass ratio
($-1<$Z/A$<+1$)
long-lived ($\tau\geq\sim$50~ns) exotic particles produced
near midrapidity ($y_{cm} \pm 0.5$) in fixed target heavy ion collisions
at a beam momentum of 11.5 GeV/c per nucleon [12,13,14,15].
At this momentum, the beam rapidity is $\sim$3.2 in the lab frame.
For this purpose, E864 is a non-focusing spectrometer with an open geometry
configuration, permitting searches over a wide range of masses
simultaneously.

\begin{figure}
\begin{center}
\epsfysize=7.5cm
\epsfbox{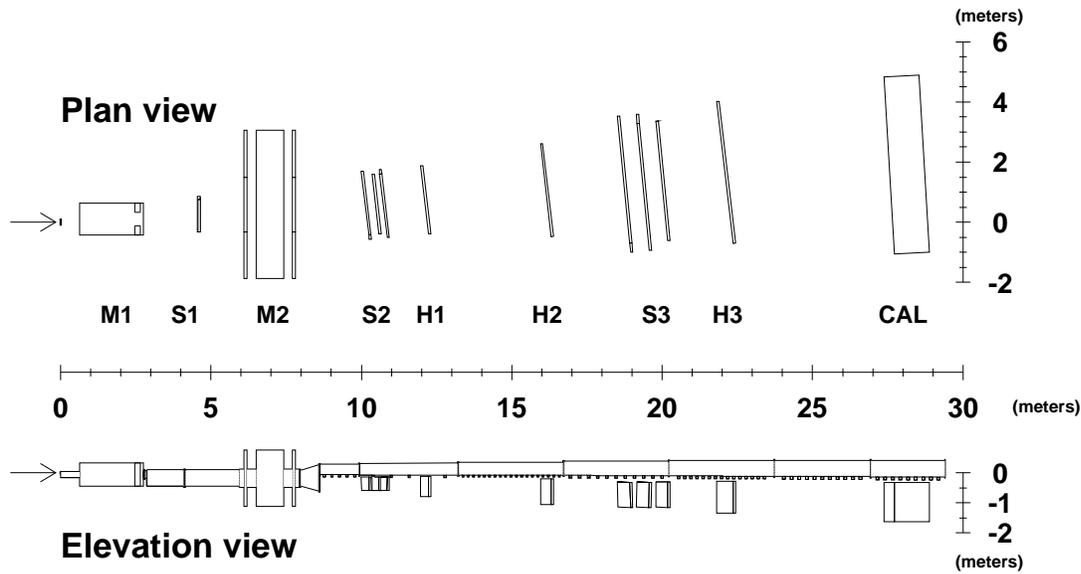}
\end{center}
\caption{Plan and elevation views of the E864 Spectrometer.}
\end{figure}

The layout of the experiment is shown in Figure~1. The spectrometer
consists of two dipole analyzing magnets (M1 and M2), 
three hodoscope scintillator planes (H1, H2, and H3),
three sets of straw tube tracking stations (S1\footnote[3]{S1 is present,
but not used in this analysis.}, S2, and S3)
with X, U, and V planes, and a hadronic calorimeter (CAL). The straw tubes
and hodoscopes are highly segmented and combine
to perform precision tracking, while the
hodoscopes also provide time-of-flight and charge measurements. This allows
particle identification with a mass resolution of $\sim$3\% at high mass.
A redundant measurement of the mass is provided by the calorimeter, which
is comprised of an array of 58 $\times$ 13 lead - scintillating fiber
towers~[16]. Energy and time resolutions
(${\Delta}E/\sqrt{E}~=~0.344/\sqrt{E}~+~0.035; \sigma_t~\approx~400$~ps)
of the calorimeter are sufficient to reject candidates from
scattered particles which yield erroneous tracking momenta.
Uninteracted beam and beam fragments are
transported above the experiment apparatus in vacuum.

The data set used in this analysis is taken from the 1996-1997 run of the
experiment using a Au beam at 11.5 GeV/c per nucleon on a 60\% interaction
length Pt target. The analyzing magnets are set to -0.75T to sweep positively
charged particles out of the spectrometer, and improve acceptance for negative
particles. A multiplicity detector located near the target [17] provides
a measure of centrality [18] and is used with counters that define a good
beam in the level I event selection trigger. The centrality requirement
is imposed from the expectation that strangelets are formed only in the
high density conditions found in highly central collisions.

The fact that the calorimeter can provide a mass measurement by itself is
utilized for a level II, Late-Energy trigger (LET).
The energy and time measurements of individual
calorimeter towers are digitized and fed into electronic, two-dimensional
lookup tables [19]. The lookup tables are programmed to provide a trigger
when an energy and time corresponding to a slow, massive particle are read out
from the calorimeter. The LET allows the experiment
to record only events in which high mass candidates are detected. In this
particular data set, only one
in $\sim$77 events which pass the level I trigger
are accepted by the level II trigger. The nearly 200 million events recorded
therefore represent approximately 14 billion sampled central collisions.

Further analysis of the data set includes requirements on the quality of
the linear track fits through the apparatus ($\chi^2$),
and an upper limit on rapidity
(the mass resolution $\sigma_m/m$ scales with $\gamma^2$ while scattering
background is worst at high rapidity). A clean corresponding calorimeter
shower with reasonable mass (energy and time) agreement is required for
consideration as a strangelet candidate [20].

\section{Results}

\subsection{Strangelets}

The spectra of charge $Z=-1$ and $Z=-2$ candidates are shown Figure~2
as a function of tracking and calorimeter masses.
A significant background exists at high tracking masses from multiple
scattering and neutron-proton charge exchange interactions downstream of
the first analyzing magnet. These are identified as having low mass
by the calorimeter and are not considered candidates. In particular,
the background in the charge $Z=-2$ spectrum yields calorimeter masses
consistent with scattered $^3$He particles. Further work is underway to
understand the sources of background at low calorimeter masses. No
high mass candidates are observed.

\begin{figure}
\begin{center}
\epsfysize=7.5cm
\epsfbox{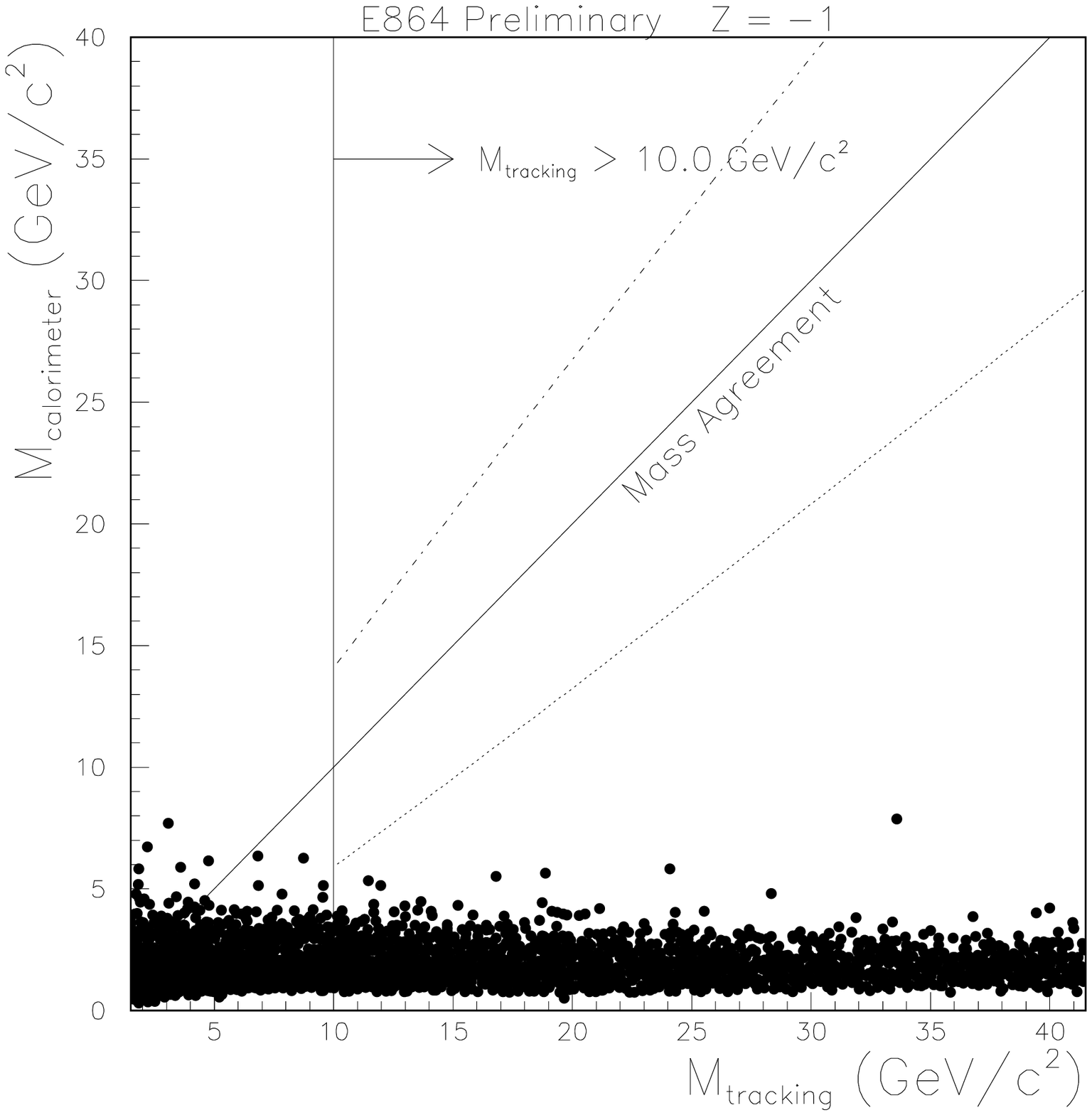}
\epsfysize=7.5cm
\epsfbox{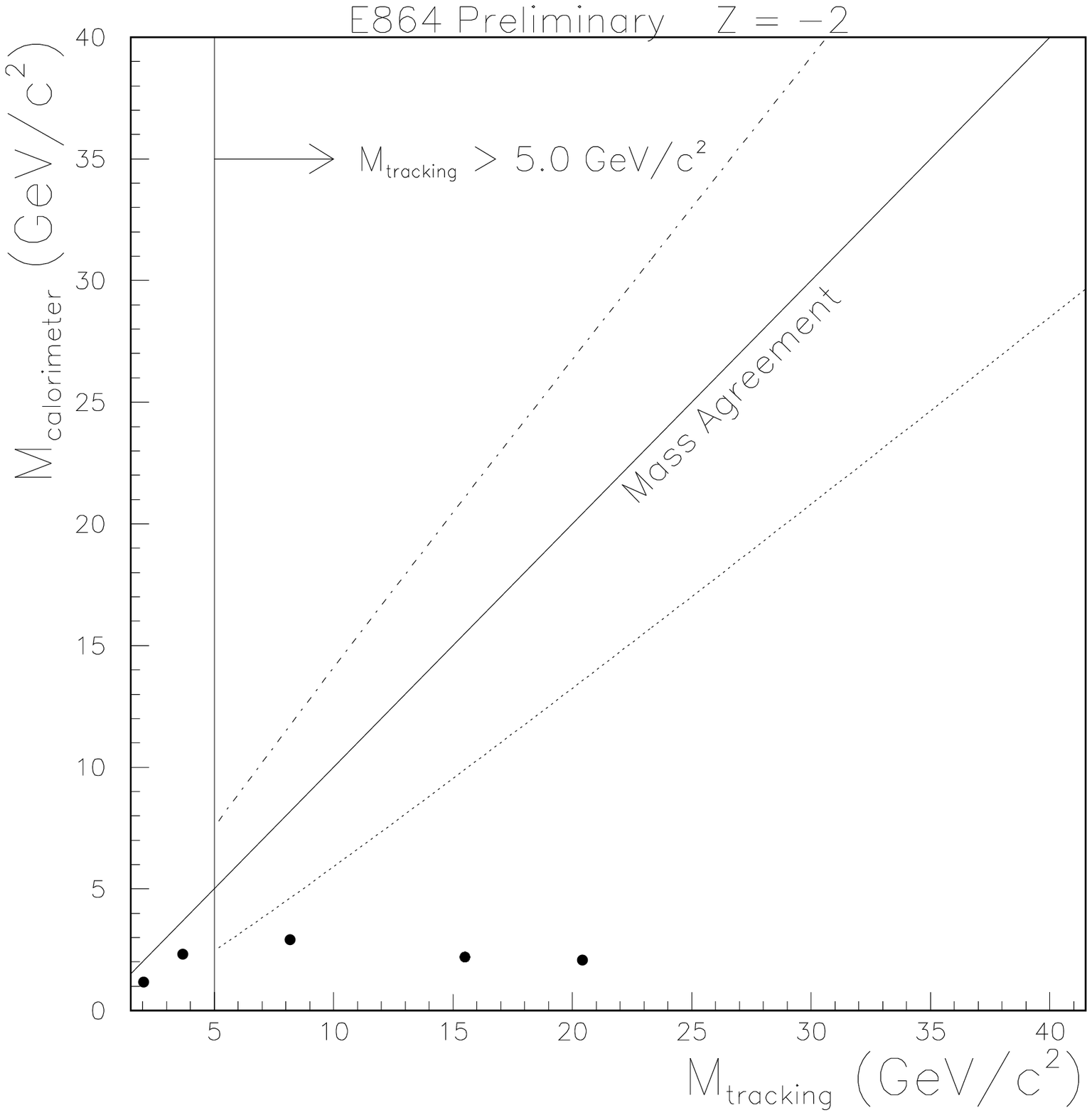}
\vskip -0.3cm
\end{center}
\caption{Spectrum of strangelet candidates for charges $Z=-1$ and $Z=-2$.
Calorimeter mass agreement of $\pm2\sigma$ curves are also shown.}
\end{figure}

In order to calculate production limits, a production model must be
assumed to determine acceptance. Two models are used assuming the
following form:
\begin{eqnarray}
\label{eq:prod_model}
\frac{d^2N}{dy \; dp_t} &\propto& p_t e^{-\frac{2 p_t}{<p_t>}}
 e^{-\frac{\left( y - y_{mid} \right)^2}{2 \sigma_y^2}}
\end{eqnarray}
where $<{p_t}> = 0.6\sqrt{A}$GeV/c is the mean transverse momentum and
$y_{mid} = 1.6$ is midrapidity. In models I and II $\sigma_y = 0.5$
and $\sigma_y = 0.5/\sqrt{A}$ respectively. However, the large acceptance of
the E864 spectrometer makes the results largely mass-~and model-independent.

\begin{figure}
\begin{center}
\epsfysize=9cm
\epsfbox{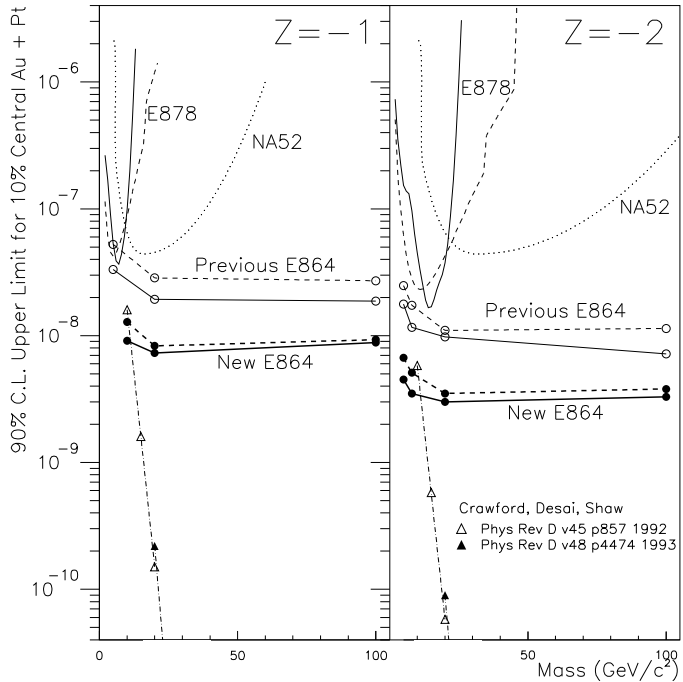}
\end{center}
\caption{90\% C.L. limits for production of charge $Z=-1$ and $Z=-2$
strangelets in 10\% central heavy ion collisions. Dashed lines represent
limits using production model~I; solid lines represent model~II. Previous
E864 limits are from the 1995 run of the experiment. NA52 results are shown
using a model similar to model~I, but are for 158 GeV/c per nucleon Pb~+~Pb
collisions~[22]. Also shown are the predictions of a QGP production
model~[23].}
\vskip -1cm
\end{figure}

Figure~3 shows the expected
upper limits calculated at the 90\% Confidence Level (C.L.)
for production of charge $Z=-1$ and $Z=-2$ strangelets in 10\% central
heavy ion collisions. These represent the best limits in the world
to date [21] and are about (averaged over all masses)
$1 \times 10^{-8}$ and $4 \times 10^{-9}$ for charge $Z=-1$ and $Z=-2$
respectively.

\subsection{QGP}

While the strangelet limits cannot tell us whether a QGP is
ever formed in heavy ion collisions, they can tell us about a particular
kind of QGP: that which decays into strangelets. The probability that
such a QGP is formed can be considered a branching fraction for the result
of a central heavy ion collision, while the probability that such a QGP
decays into a charge $Z=-1$ or $Z=-2$ strangelet
can also be considered a branching
fraction. An upper limit is then calculated for each branching fraction
as a function of the other, as shown in Figure~4.

\begin{figure}
\begin{center}
\epsfysize=7cm
\epsfbox{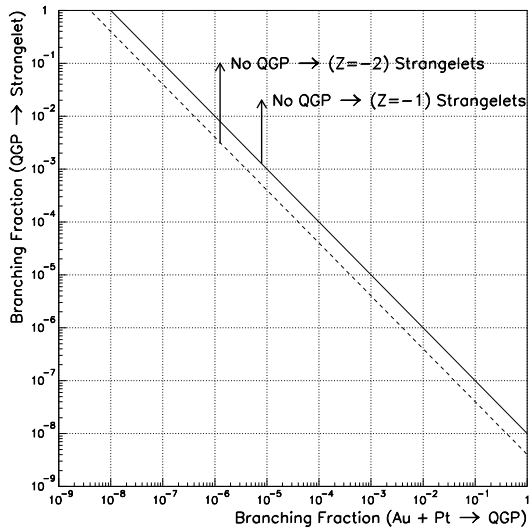}
\end{center}
\caption{90\% C.L. upper limits on the branching fractions of heavy ion
collisions into a strangelet-producing QGP, and subsequent decay into
charge $Z=-1$ and $Z=-2$ strangelets.}
\end{figure}

\section{Conclusion and Future Prospects}

E864 has found
no clear evidence for negatively charged strangelets in central Au~+~Pt
collisions at 11.5 GeV/c per nucleon. We have set 90\% C.L. upper limits
of ${\sim}1 \times 10^{-8}$ and ${\sim}4 \times 10^{-9}$ per central
collision for production of
charge $Z=-1$ and $Z=-2$ strangelets with proper lifetimes $\tau\geq\sim$50~ns
respectively. While it is difficult to make any statements about coalescence
and thermal model production mechanisms, upper limits have been calculated on
the formation of a QGP which distills into such strangelets.

Further work on this front includes looking for strangelets which have
charge $Z\leq-3$ in the data set, and conducting a search for strangelets
which may interact as antimatter in the E864 calorimeter. The final
sensitivities from E864 will be further enhanced by combining the results
of this data set (taken at -0.75T) with those from the +1.5T data set
taken during the same running period. These limits will likely be the
ultimate results for negatively charged strangelet searches at the AGS [21].

\section{Acknowledgements}

The E864 collaboration wishes to thank the AGS staff for
their support. This work was supported by the Department
of Energy's High Energy and Nuclear Physics Divisions, the U.S. National
Science Foundation, and the Instituto di Fisica Nucleara of Italy.

\end{document}